# Interpretable 3D Multi-Modal Residual Convolutional Neural Network for Mild Traumatic Brain Injury Diagnosis.

Hanem Ellethy [1][0000-0003-2791-9087], Viktor Vegh [2,3,4][0000-0002-4310-1421], and Shekhar S. Chandra [1][0000-0001-6544-900]

[1] School of Electrical Engineering and Computer Science, University of Queensland, QLD, Australia.
[2] The Centre for Advanced Imaging, University of Queensland, QLD, Australia.
[3] Australian Institute for Bioengineering and Nanotechnology, University of Queensland, QLD, Australia.
[4] ARC Training Centre for Innovation in Biomedical Imaging Technology, QLD, Australia.

`h.elwaseif@uq.edu.au`

**Abstract.** Mild Traumatic Brain Injury (mTBI) is a significant public health challenge due to its high prevalence and potential for long-term health effects. Despite Computed Tomography (CT) being the standard diagnostic tool for mTBI, it often yields normal results in mTBI patients despite symptomatic evidence. This fact underscores the complexity of accurate diagnosis. In this study, we introduce an interpretable 3D Multi-Modal Residual Convolutional Neural Network (MRCNN) for mTBI diagnostic model enhanced with Occlusion Sensitivity Maps (OSM). Our MRCNN model exhibits promising performance in mTBI diagnosis, demonstrating an average accuracy of 82.4%, sensitivity of 82.6%, and specificity of 81.6%, as validated by a five-fold cross-validation process. Notably, in comparison to the CT-based Residual Convolutional Neural Network (RCNN) model, the MRCNN shows an improvement of 4.4% in specificity and 9.0% in accuracy. We show that the OSM offers superior data-driven insights into CT images compared to the Grad-CAM approach. These results highlight the efficacy of the proposed multi-modal model in enhancing the diagnostic precision of mTBI.

## 1 Introduction

Mild Traumatic Brain Injury (mTBI), representing the majority of annual global traumatic brain injuries (TBI), leads to serious and enduring consequence [1]. Despite its prevalence, the diagnosis of mTBI remains a substantial challenge [2]. Traditionally, mTBI diagnosis relies on subjective clinical evaluations, including symptoms and cognitive tests, which often lack consistency and accuracy [3]. Computed Tomography (CT), despite its limitations in detecting subtle or non-existent structural changes in mTBI cases [4], is the standard neuroimaging test due to its availability and speed in assessing intracranial lesions [5].

Machine learning, specifically deep learning (DL), has shown promise in medical imaging diagnosis and condition prediction [6], [7]. However, its application in diagnosing mTBI using CT data remains rare [8]. DL algorithms present a valuable opportunity to enhance mTBI diagnosis by extracting subtle, clinically meaningful information from complex CT data, which may not readily be apparent to human observers [9]. In our previous study [10], Artificial Neural Network (ANN) and Random Forest (RF) approaches demonstrated remarkable performance in diagnosing mTBI using clinical and image interpreted data. However, employing imaging data directly to build an automated mTBI diagnostic model promises to be more feasible, effective, and time saving. Radiologists screen and report mTBI-related findings from a CT scan, such as hemorrhage, hematomas, swelling, skull fracture, and any obvious tissue damage [11]. While conventional computer vision techniques such as the scale-invariant feature transform algorithm can perform feature detection/extraction [12], DL models have been shown to be more effective for extracting highly complex and task-specific features from images [13]. Therefore, an automated CT diagnosis system for mTBI could substantially improve diagnostic speed, decision-making, and resource efficiency, potentially reducing morbidity and mortality rates.

Several studies have applied DL techniques successfully, particularly using Convolutional Neural Networks (CNN), to construct TBI diagnostic models based on CT scans [14], [15]. These studies underline the potential of CNNs to segment and/or identify moderate and severe TBI hemorrhage lesions on CT images. However, the detection and interpretation of mTBI lesions on CT scans remain unclear and controversial. While many studies have used ML and DL for mTBI diagnosis using a range of data sources [16]–[18], these researches have not fully extended to the utilization of CT data. Thus, the potential of CT data for this purpose remains largely untapped.

Here, we introduce a novel Multi-Modal Residual Convolutional Neural Network (MRCNN) equipped with Occlusion Sensitivity Maps (OSM) for the diagnosis of mTBI, expanding the application of DL techniques in this field. Our work also provides an innovative comparison between OSM and Gradient-weighted Class Activation Mapping (Grad-CAM), offering new perspectives on their usage for data-driven insights into CT images.

The main contribution of this study can be summarized as follows:

— We developed a Multi-Modal Residual Convolutional Neural Network (MRCNN) enhanced with occlusion sensitivity. This represents a novel approach in the diagnosis of mTBI, effectively integrating CT imaging and clinical data to improve diagnostic precision.
— We introduce the application of OSM in the context of mTBI diagnosis, offering superior data-driven insights into CT images compared to the widely used Gradient-weighted Class Activation Mapping (Grad-CAM).
— We demonstrate the efficacy of integrating multiple modalities into the model to enhance diagnostic accuracy and precision. The effectiveness of this multi-modal approach is substantiated by the tangible improvements in specificity and accuracy compared to the proposed CT-based Residual Convolutional Neural Network (RCNN) model.

## 2 Methodology

We explored different DL architectures to extract and learn the spatial features of mTBI from 3D CT scans. Methods were selected based on their previously demonstrated ability to work with complex patterns within images. In the initial stage, clinical features were used to train the classifier employed in our previous study[10]. This was done to ensure that the clinical data used for the model's training aligned with evidence-based research pertinent to mTBI diagnosis. The selected features are the relevant clinical features that correspond with the feature ranking from our prior study of the pediatric emergency care applied research network (PECARN) dataset [10] for mTBI diagnosis. Following this, we applied DL models for mTBI diagnosis from CT imaging data.

### 2.1 Data preparation

TRACK-TBI pilot study dataset [19] was employed in our research to develop deep learning-based diagnostic models for mTBI utilizing CT imaging data, and to assess the performance and integration capabilities of these models. TRACK-TBI pilot study is a prospective multi-center study that enrolled 600 patients older than seven years of age and having a positive TBI diagnosis. All patients had received a head CT scan while in the emergency department within 24 h of head injury. Data were collected in three level I trauma centers (University of California San Francisco, University Medical Centre Brackenridge, University of Pittsburgh Medical Centre) and one rehabilitation center (Mount Sinai Rehabilitation Centre) in the US between April 2010 and June 2012 [19]. We obtained access privileges to the data repository (i.e., FITBIR) from the Data Access and Quality (DAQ) committee. Our study using this data was approved by the human research ethics at the University of Queensland (no. 2020002583). The patients who had undergone non-contrasted CT scans within 48 hours following closed head injury and were assigned Glasgow Coma Scale (GCS) ≥ 13 were selected for our study. A total of 296 de-identified records were sourced from the TRACK-TBI public dataset.

### 2.2 Dataset pre-processing

We pre-processed CT images to ensure high-quality, consistent data were fed into our DL models. Initially, the dcm2nii tool [20] was used to convert the CT images from their original format to the Neuroimaging Informatics Technology Initiative (NIfTI) format, incorporating field-of-view alignment corrections to have spatial consistency across all images. Next, the appropriate brain window (window width = 80 HU and width level = 40 HU) [21] was applied to standardize CT images' dynamic range and to visually enhance potential mTBI lesions. Next, the Spline Interpolated Zoom (SIZ) resampling volume algorithm was used to create isotropic resolution images [22] resulting in 128×128×64 matrix size, and 1mm3 resolution images. This resizing and uniformization, aimed to reduce computational complexity and optimize GPU usage, utilized z-axis interpolation for uniform volume representation. This approach effectively captured data from multiple slices and ensured robust performance in line with GPU memory requirements [23]. Lastly, unnecessary background information in images was removed, e.g., the scanner bed, to focus the model solely on the brain. Fig. 1 illustrates the pre-processing steps, and the Grad-CAM analysis.

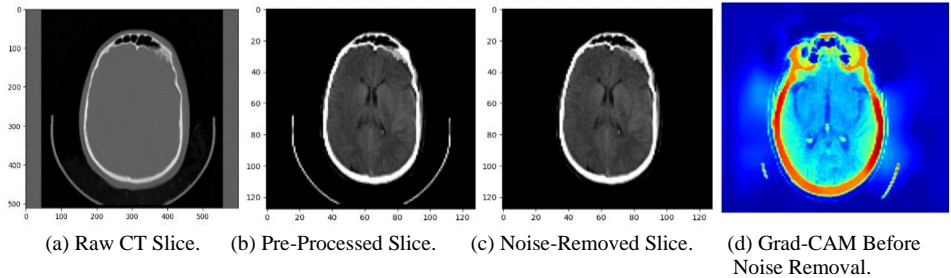

(a) Raw CT Slice.  (b) Pre-Processed Slice.  (c) Noise-Removed Slice.  (d) Grad-CAM Before Noise Removal.

**Fig. 1.** The image pre-processing steps applied to a CT slice and the impact of noise removal on the model's decision-making process. (a) An original, raw CT slice before any pre-processing steps are applied, (b) The same CT slice after it has undergone initial pre-processing steps, including standardizing the dynamic range for brain anatomical structures, (c) The CT slice has been further processed to remove high-intensity noise artifacts such as the scanner bed, and (d) illustrates how the model was significantly influenced by the high-intensity noise artifacts before their removal, as visualized by the Gradient-weighted Class Activation Mapping (Grad-CAM)

## 2.3    RCNN model

In our initial attempt to construct an mTBI diagnostic DL model, we implemented a CNN model [22]. This was followed by trials with several other architectures, encompassing variations of 3D EfficientNet, ResNet, and DenseNet [24]. The unique challenge in developing mTBI diagnostic models arises from its inherent heterogeneity, with factors such as varied injury mechanisms and differing patient symptoms adding to the complexity. Consequently, the models tested did not yield satisfactory performance. Amidst these complexities, the model architecture depicted in Fig. 2 demonstrated superior performance. As such, we adopted a tuned Residual CNN as our primary CT diagnostic model for mTBI. The model involves six residual blocks, each comprised of three convolutional blocks and a 1×1 convolution layer. Each convolutional block integrates a 3d convolutional layer, an instance normalization layer, and a parametric rectified linear unit (PReLU) activation function. The output from these six residual blocks is then flattened and passed to three fully connected layers, with a sigmoid activation function in the output layer serving as the classifier output.

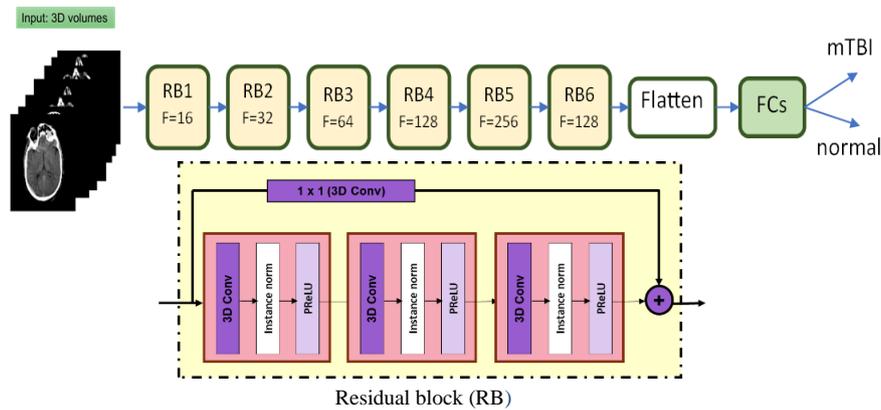

**Fig. 2.** Image-based mTBI diagnostic model architecture.
FCs: fully connected layers

### 2.4 MRCNN

Multi-modal refers to the use of both clinical information in conjunction with CT scans. We chose a multi-modal approach that leverages both clinical data and 3D CT scans to enhance the diagnostic performance of our model. This clinical data encompasses patient demographics, medical history, and symptom profiles, as depicted in Fig. 3. We aimed to improve the classification accuracy by providing the model with comprehensive patient information by integrating the diversity of data sources into a DL framework. Our model, MRCNN, is based on the foundational architecture of RCNN. It includes an additional branch for clinical data and an infusion layer designed to merge the learned CT features with the clinical features. This combined data is then processed through an output layer equipped with a sigmoid activation function.

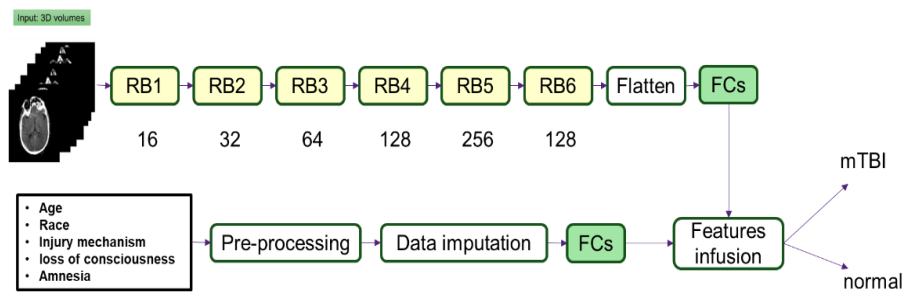

**Fig. 3.** MRCNN model architecture.

## 2.5 Implementation Details

Rectified Adaptive moment estimation (RAdam) [25] is adopted as the training optimizer with an initial learning rate of 0.0005, in conjunction with a cosine annealing scheduler [26]. The initial choice of hyperparameters, such as the number of filters used, was initially guided by insights gained from other studies [22], [24], and subsequently fine-tuned through experimental trials, with the aim of optimizing validation accuracy.

Furthermore, image intensity normalization and online augmentation techniques such as flipping, rotating, and zooming were employed during the training phase. Specifically, random rotations were applied within a range of ±10 degrees along the x, y, and z axes. Random zooming was implemented within a range of 0.9 to 1.2 times the original size. Also, vertical flipping was performed on the images along the y-axis. These strategies were adopted to address the challenges related to our limited dataset size, introduce additional variation, and to improve model generalizability [27].

We employed five-fold cross-validation as a strategy to evaluate our models' performance, individually assessing accuracy, sensitivity, and specificity using 95% confidence intervals (CI) as done previously [28]. This evaluation process helps in identifying the strengths and limitations of the model, thereby facilitating informed comparisons. Moreover, we enhanced visual understanding of the model's decision-making process using OSM [29], to provide an intuitive way for data interpretation for mTBI diagnosis compared to Grad-CAM [30]. The occlusion sensitivity map is a method to understand a model's decision-making process. It involves systematically occluding different parts of an input image and observing how these alterations affect the model's output. By doing so, it helps to identify the regions in the input image that the model considers critical for decision making. This approach adds depth to model evaluation by highlighting how image regions influence the decisions made by the model.

## 3 RESULTS AND DISCUSSION

### 3.1 Metrics Evaluation

The RCNN model achieved promising results in diagnosing mTBI using CT images only with an average sensitivity of 82.6% and specificity of 72.7 % as reported in Table 1 with its mean and standard deviation. The features corresponding to those ranked in the PECARN-based study [10] included age, race, Glasgow Coma Scale (GCS), injury mechanism, loss of consciousness, and amnesia. However, GCS was excluded as there was not a sufficient level of variation across patients. Using only these five clinical features to train the classifier [10], it achieved an average accuracy of 57.1%, a sensitivity of 69.1%, and a specificity of 50.0% for diagnosing mTBI using the clinical data only.

Notably, the MRCNN model exhibited an improvement in performance over RCNN with an average accuracy of 82.4% and specificity of 81.6%. With the integration of clinical data in the RCNN framework, accuracy and specificity increased by 4.4% and 9.0%, respectively. It does appear that the inclusion of clinical data offers a more comprehensive mTBI patient representation, extending beyond what can be seen in CT scans alone.

**Table 1.** Evaluation metrics of the five-fold cross-validation.

| Metric | Clinical | Imaging | Multi-modal |
|---|---|---|---|
|  | ANN [10] | RCNN | MRCNN |
| **Accuracy (μ ± α in %)** | 57.1 ± 7.8 | 78.1 ± 6.6 ($p_{ANN}$ = 0.002) | 82.4 ± 1.7 ($p_{RCNN}$ = 0.024) |
| **Sensitivity (μ ± α in %)** | 69.1 ± 14.6 | 82.8 ± 5.1 ($p_{ANN}$ = 0.061) | 82.6 ± 5.1 ($p_{RCNN}$ = 0.294) |
| **Specificity (μ ± α in %)** | 50.0 ± 10.7 | 72.7 ± 8.7 ($p_{ANN}$ = 0.006) | 81.6 ± 5.4 ($p_{RCNN}$ = 0.022) |

ANN: Artificial Neural Network, μ: Mean, α: Standard Deviation. Statistically significant increase in the mean metric was achieved in all cases for $p < 0.05$, except for sensitivity between RCNN and MRCNN.

In addition to our primary metrics, we assessed the overall performance of our model using the Area Under the Curve (AUC) of the Receiver Operating Characteristic (ROC) curve. As illustrated in Fig. 4, the ROC curve graphically represents the trade-off between the true positive rate and false positive rate of the MRCNN model across varying discrimination thresholds. The obtained AUC score of 0.95 signifies an excellent ability of our model to differentiate between mTBI and normal CT scans.

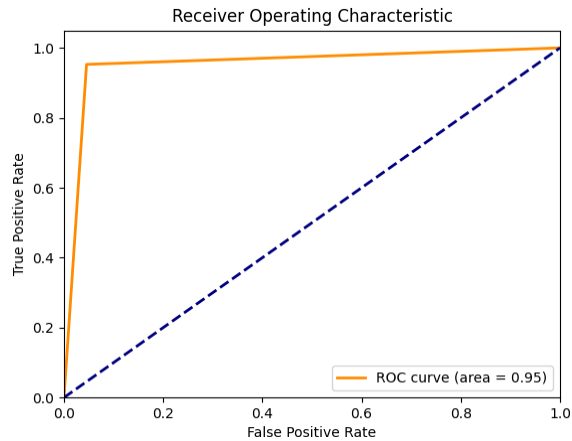

**Fig. 4.** The Area Under the Curve of the Receiver Operating Characteristic (AUC) curve of the MRCNN model.

Our results have profound implications as this may be one of the first attempts to employ DL in diagnosing mTBI using CT images. The diagnosis of mTBI is an exceptionally challenging task due to its subtle nature and the limited sensitivity of CT scans. However, the promising performance of the MRCNN model demonstrates the potential of DL to improve mTBI diagnosis using CT imaging and clinical data. It suggests that DL can potentially extract clinically relevant information from these scans that might otherwise be overlooked. Therefore, this work paves the way for further research on the application of DL models in leveraging CT images for mTBI diagnosis, offering new avenues for early detection and treatment [3].

### 3.2 Visual Assessment

A vital question that emerges with medical diagnostic models is whether the model truly identifies and relies on meaningful features, or if it is simply capitalizing on coincidental patterns. This underscores the importance of exploring visual explanations for the decision-making process involved in the model. For our RCNN model, we achieved this through the application of Grad-CAM and occlusion sensitivity, as depicted in Fig. 5.

While Grad-CAM is a well-established technique for exploring the decisions of CNNs within the scope of computer vision tasks, it does have certain limitations [31]. In our case, it primarily highlighted the skull because of its high intensity values, with less emphasis on the brain. This limitation is evident in the third column of Fig. 5, where the brain regions are under-represented compared to the skull and values with high intensities.

In contrast, OSM, generated by systematically blocking various parts of the input image, yields a more comprehensive, nuanced understanding. As the model output changes due to the occlusion of certain regions, a region's importance in the decision-making process is scored [32]. The technique thus provides a more comprehensive visualization of scans, focusing on both the skull and the interior of the brain.

The comparative efficiency of occlusion sensitivity over Grad-CAM becomes clear in Fig. 5. While Grad-CAM primarily focuses on the skull area, occlusion sensitivity highlights the regions that substantially influence the decisions made by the model. The blue areas on the OSM represent the regions that, when occluded, have the most significant impact on the output, indicating their relevance to the model.

In the correctly classified mTBI CT scan, the occlusion sensitivity map highlights not only a part of the skull but also a significant portion of the brain. This indicates that the model correctly associates certain features within these regions with the diagnosis of mTBI. This contrasts with the Grad-CAM results, which primarily focus on the high-intensity areas of the skull.

For the correctly classified normal CT scan, there's a more significant focus on the skull and the brain areas adjacent to the skull in the occlusion sensitivity map. This suggests that the model correctly identifies these areas as important for a normal diagnosis, indicating a distinct difference from the features it associates with mTBI.

In the misclassified CT scan, attention is given to different areas, both in the skull and the brain, in the occlusion sensitivity map. It suggests that the model might be incorrectly focusing on these areas or missing capturing information on other important features, leading to misclassification. Understanding these less significant areas in terms of decision-making by the model could provide valuable insights for improving model performance.

Therefore, the OSM can reveal areas that substantially alter the output of the model when occluded, capturing vital details that could potentially be overlooked by Grad-CAM. These insights from OSM can greatly enhance the clinical utility of our model by significantly contributing to the understanding and diagnosis of mTBI.

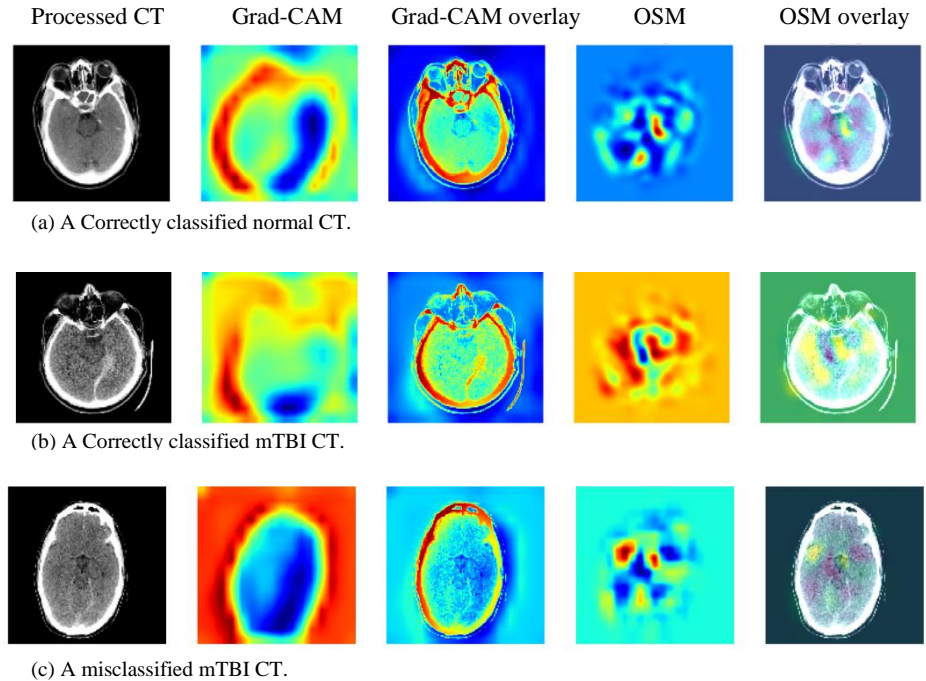

**Fig. 5.** The visual explanation of the model. From left to right, the columns respectively represent the processed CT scan, the Grad-CAM visualization, the overlay of the Grad-CAM and the processed CT scan, the occlusion sensitivity maps, and finally the overlay of the occlusion sensitivity maps with the processed CT scan.

## 4 Conclusions and Future Work

We introduced a CT-based interpretable Multi-modal Residual Convolutional Neural Network (MRCNN) model for mild traumatic brain injury (mTBI) diagnosis enhanced with Occlusion Sensitivity Maps (OSM). Remarkably, the MRCNN model improved the average specificity by an additional 4.4% and accuracy of 9.0%, indicating that the integration of clinical data with CT imaging offers a more thorough patient representation and improves the precision achieved by the MRCNN. Furthermore, the use of OSM for visual explanation adds significant value to model interpretability. This technique aids in understanding the regions of interest that the model utilizes for mTBI diagnosis, resulting in improved transparency and creating trust in the model output.

Despite our interesting and promising results for mTBI, the diagnostic model has several limitations that need to be addressed. It requires high-quality, accurately labelled CT scans for training. The performance of the model can be significantly impacted by the quality and diversity of the training data. Therefore, collecting and curating such datasets is challenging and requires significant effort. Future work should aim to address these limitations and further improve the model. Potential avenues could include the addition of other clinical data to create a more comprehensive diagnostic tool or exploring different neural network architectures achieving even better performance than presented here.

## Acknowledgement

NITRC, NITRC-IR, and NITRC-CE have been funded in whole or in part with Federal funds from the Department of Health and Human Services, National Institute of Biomedical Imaging and Bioengineering, the National Institute of Neurological Disorders and Stroke, under the following NIH grants: 1R43NS074540, 2R44NS074540, and 1U24EB023398 and previously GSA Contract No. GS-00F-0034P, Order Number HHSN268200100090U. Moreover, we would like to acknowledge the principal investigators of the TRACK TBI Pilot research program, the sub-investigators and research teams that contributed to TRACK TBI Pilot, and the patients who participated.